\documentclass[12pt]{amsart}
\usepackage{latexsym, amsmath, amssymb,amsthm,amsopn,amsfonts,mathrsfs,tikz}
\usepackage{stmaryrd}
\usepackage{version}
\usepackage{epsfig,graphics,color,graphicx,graphpap,dsfont,eufrak}
\usepackage{amssymb}
\usepackage{geometry}
\usepackage{tikz}
\usetikzlibrary{snakes}

\usepackage{graphicx}
\usepackage{subfigure}

\ifx\pdfoutput\undefined
  \DeclareGraphicsExtensions{.eps}
\else
  \ifx\pdfoutput\relax
    \DeclareGraphicsExtensions{.eps}
  \else
    \ifnum\pdfoutput>0
      \DeclareGraphicsExtensions{.pdf}
    \else
      \DeclareGraphicsExtensions{.eps}
    \fi
  \fi
\fi

\newcommand{\bse}{\begin{subequations}}
\newcommand{\ese}{\end{subequations}}

\DeclareFontFamily{OT1}{pzc}{}
\DeclareFontShape{OT1}{pzc}{m}{it}{<-> s * [1.10] pzcmi7t}{}
\DeclareMathAlphabet{\mathpzc}{OT1}{pzc}{m}{it}

\theoremstyle{plain}

\theoremstyle{definition}

\numberwithin{equation}{section}

\def\squarebox#1{\hbox to #1{\hfill\vbox to #1{\vfill}}}

\usepackage{amsxtra}

\usepackage{hyperref}

\renewcommand{\k}{\kappa}
\newcommand{\ga}{\gamma}

\newcommand{\ra}{\rightarrow}
\newcommand{\al}{\alpha}
\newcommand{\be}{\begin{equation}}
\newcommand{\ee}{\end{equation}}

\newcommand{\pa}{\partial}

\begin{document}

\title[Enrichment paradox and applications]{Enrichment paradox and applications}

\author{Z. C. Feng}
\address{Department of Mechanical and Aerospace Engineering, 
University of Missouri, Columbia, MO 65211}
\email{fengf@missouri.edu}

\author{Y. Charles Li}
\address{Y. Charles Li, Department of Mathematics, University of Missouri, 
Columbia, MO 65211, USA}
\email{liyan@missouri.edu}
\urladdr{http://faculty.missouri.edu/~liyan}

\begin{abstract}
We introduced a more general predator-prey model to analyze the paradox of enrichment. We hope the results obtained for the model can guide us on identifying real field paradox of enrichment. 
\end{abstract}

\maketitle

\section{Introduction}

Human population (especially in the third world countries) has dramatically increased due to enrichment of food resource such as high yield grains and GMOs. The crucial question is whether or not such a huge human population is sustainable. What will happen if a worldwide famine takes place? Will the human population go extinct or suffer regional eradication? One can view the human-food 
system as a predator-prey system. The predator-prey dynamics is everywhere in ecological systems. Intuitively speaking, when the food supply of the prey is enriched, both the prey and the predator populations should rise. On the other hand, is it possible that 
enrichment can lead to extinction of both or one of predator and prey? If this happens, we have a paradox of enrichment. Since there is a variety of dynamics among different predator-prey systems, some predator-prey systems may have the paradox of enrichment. For instances, if an enrichment changes the intensity of predation, then extinction of both or one of predator and prey may be 
possible. If an enrichment increases the intensity of predation, then the prey together with the predator may be extinct. If an enrichment decreases the intensity of predation, then the predator may be extinct. Nevertheless, so far a real field example of the 
paradox of enrichment is still elusive. Mathematical models on the predator-prey dynamics indeed often predict the existence of the paradox of enrichment. Mathematical models can provide guidance for identifying predator-prey systems that have the 
paradox of enrichment. A general predator-prey model is the Kolmogorov model:
\[
\frac{dU}{dt} = U f(U,V), \  \frac{dV}{dt} = V g(U,V), 
\]
where $U$ is the population of prey, $V$ is the population of predator, and
\[
\frac{\pa f}{\pa V} < 0, \ \frac{\pa g}{\pa U} > 0. 
\] 
By choosing 
\[
f(U,V) = \al - \ga V,  \ g(U,V) = \k U - \mu , 
\]
one gets the simple Lotka-Volterra model:
\[
\frac{dU}{dt} = \al U - \ga UV, \  \frac{dV}{dt} = \k UV - \mu V, 
\]
where $\al$ is the birth rate of the prey, $\ga$ is the coefficient of predation, $\k$ is the coefficient of food utilization of the predator, and $\mu$ is the mortality rate of the predator. The Lotka-Volterra model does not take into account the effect of carrying 
capacity of the prey, which is the upper limit of the prey population that the food supply can support. The Holling model incorporated the effect of carrying capacity,
\[
\frac{dU}{dt} = \al U \left ( 1-\frac{U}{b} \right ) - Vf(U), \  \frac{dV}{dt} = \k V f(U) - \mu V, 
\]
where $b$ is the carrying capacity of the prey, and $f(U)$ is the functional response of Holling's type I, II and III. A type I functional response is used in the Lotka-Volterra model. A type II functional response is used in the Rosenzweig-Macarthur model,
\[
\frac{dU}{dt} = \al U \left ( 1-\frac{U}{b} \right ) - \ga \frac{U}{U+h}V, \  \frac{dV}{dt} = \k \ga \frac{U}{U+h}V - \mu V, 
\]
where $\al$ is the birth rate of the prey, $b$ is the carrying capacity of the prey, $h$ is the half-predation parameter, $\ga$ is the coefficient of predation, $\k$ is the coefficient of food utilization of the predator, and $\mu$ is the mortality rate of the 
predator. The paradox of enrichment originated from mathematical analysis on the Rosenzweig-Macarthur model. The 
crucial question is whether or not the paradox of enrichment exists in reality or it is simply a mathematical artifact of the 
model. The following Arditi-Ginzburg model does not have the paradox of enrichment,
\[
\frac{dU}{dt} = \al U \left ( 1-\frac{U}{b} \right ) - Vf(\frac{U}{V}), \  \frac{dV}{dt} = \k V f(\frac{U}{V}) - \mu V.
\]
Since there is a variety of predator-prey dynamics in reality, there is no universal mathematical model that can model all the predator-prey dynamics. In this paper, we are interested in studying the following more general model
\begin{equation}
\frac{dU}{dt} = \al U \left ( 1-\frac{U}{b} \right ) - Vf(\frac{U}{V^{\nu}}), \  \frac{dV}{dt} = \k V f(\frac{U}{V^{\nu}}) - \mu V,
\label{GM}
\end{equation}
when $\nu = 0$, the model reduces to the Holling model, when $\nu = 1$, the model reduces to the Arditi-Ginzburg model, and in general $\nu \geq 0$. The introduction of the more general form of functional response $f(\frac{U}{V^{\nu}})$ is motivated 
by the Kleiber's law in biology where we can draw the analogy of the mass of an animal to the population of predator and the 
animal's metabolic rate to the population of prey. The Kleiber's law states that an animal's metabolic rate scales to the $3/4$ power of the animal's mass.

\section{Dynamics of a more general predator-prey model}

\begin{figure}[ht] 
\centering
\includegraphics[width=5.0in,height=4.0in]{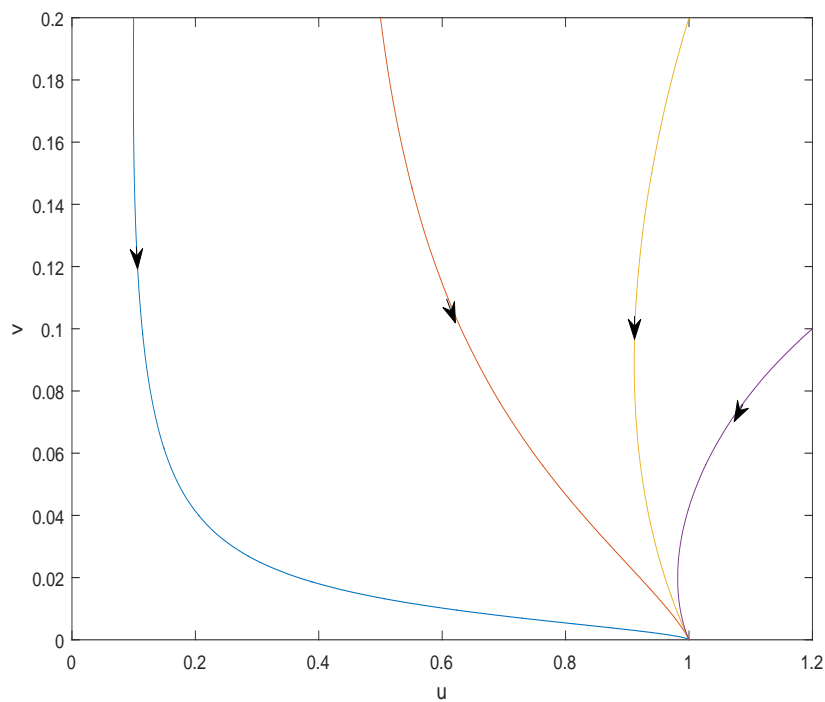}
\caption{Dynamics of (\ref{sm1})-(\ref{sm2}) when $r = 3/2, k = 1, H = 0.1, \nu = 0$.}
\label{F1}
\end{figure}

\begin{figure}[ht] 
\centering
\includegraphics[width=5.0in,height=4.0in]{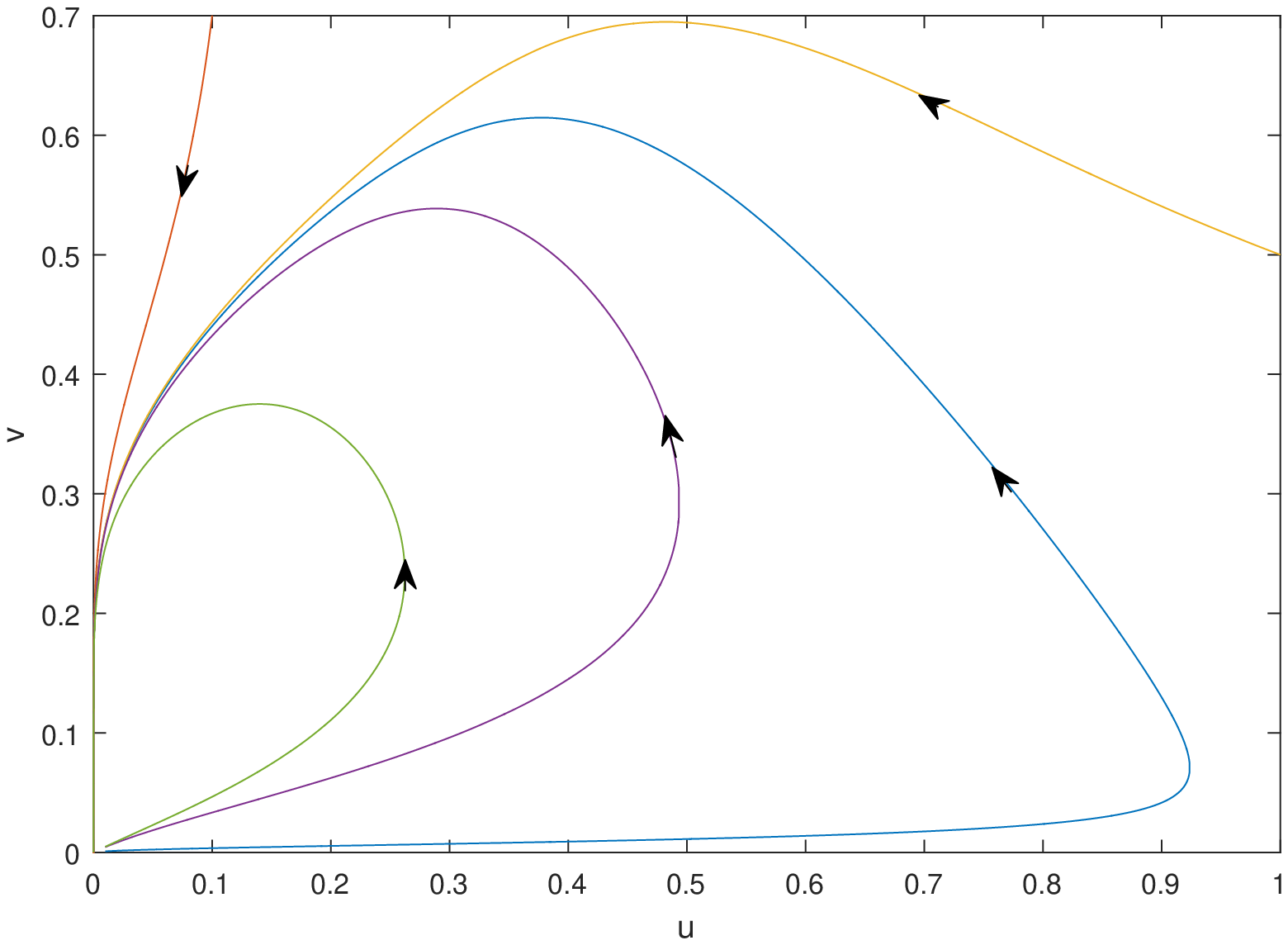}
\caption{Dynamics of (\ref{sm1})-(\ref{sm2}) when $r = 1/2, k = 1, H = 1, \nu = 2$.}
\label{F4}
\end{figure}

\begin{figure}[ht] 
\centering
\includegraphics[width=5.0in,height=4.0in]{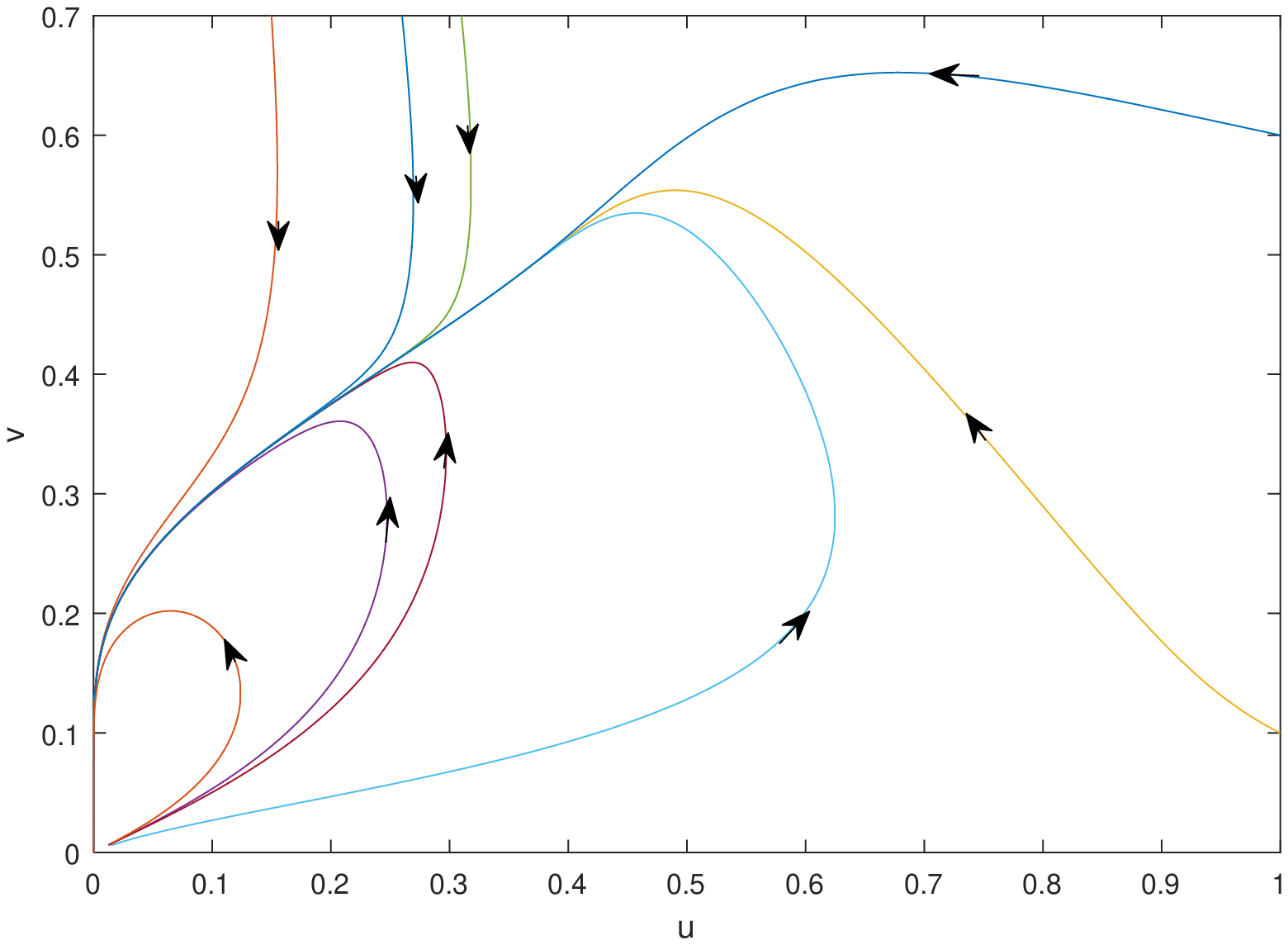}
\caption{Dynamics of (\ref{sm1})-(\ref{sm2}) when $r = 1/2, k = 1, H = 1.6, \nu = 2$.}
\label{F5c}
\end{figure}

\begin{figure}[ht] 
\centering
\includegraphics[width=5.0in,height=4.0in]{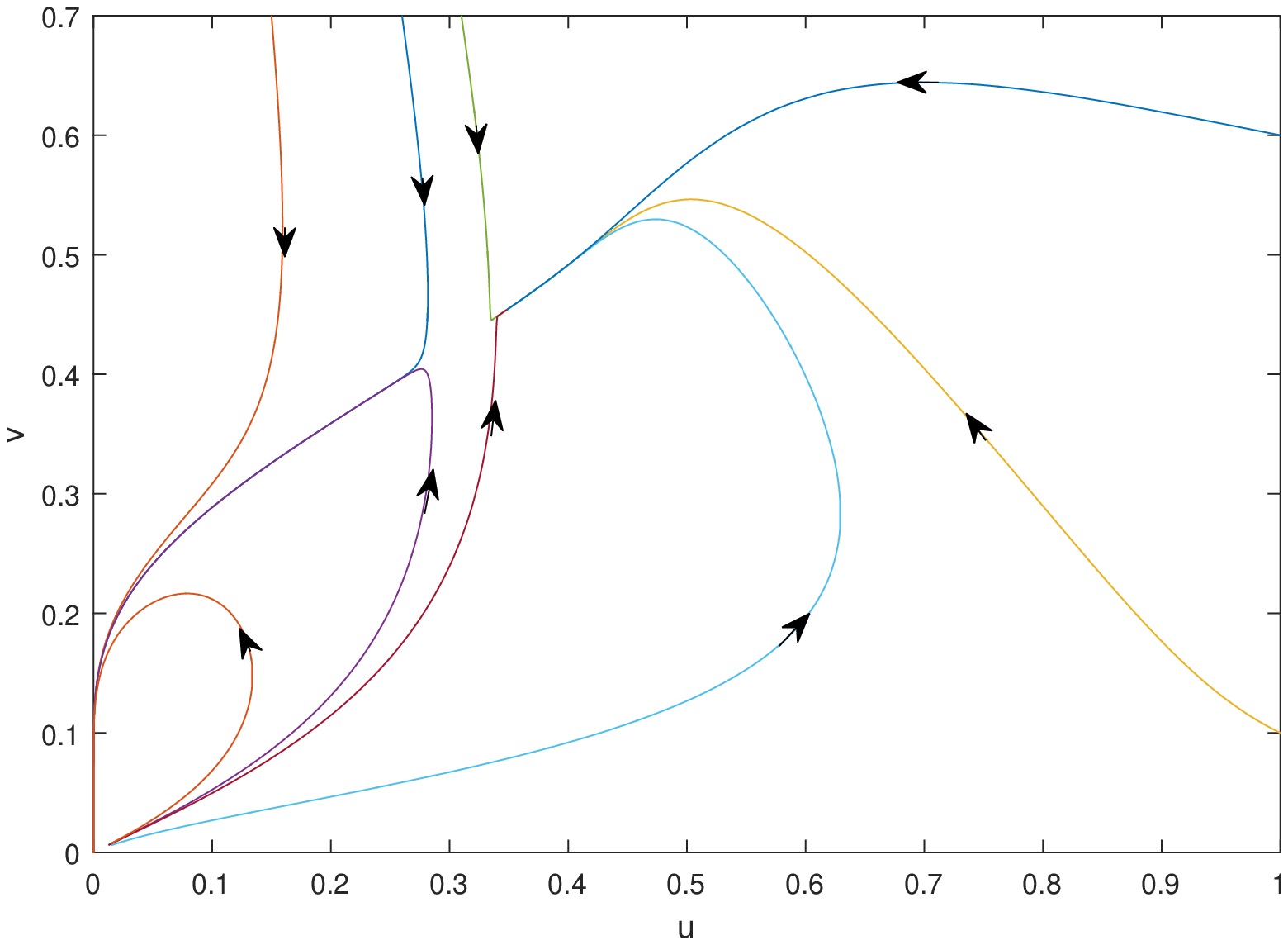}
\caption{Dynamics of (\ref{sm1})-(\ref{sm2}) when $r = 1/2, k = 1, H = 1.69, \nu = 2$.}
\label{F5b}
\end{figure}

\begin{figure}[ht] 
\centering
\includegraphics[width=5.0in,height=4.0in]{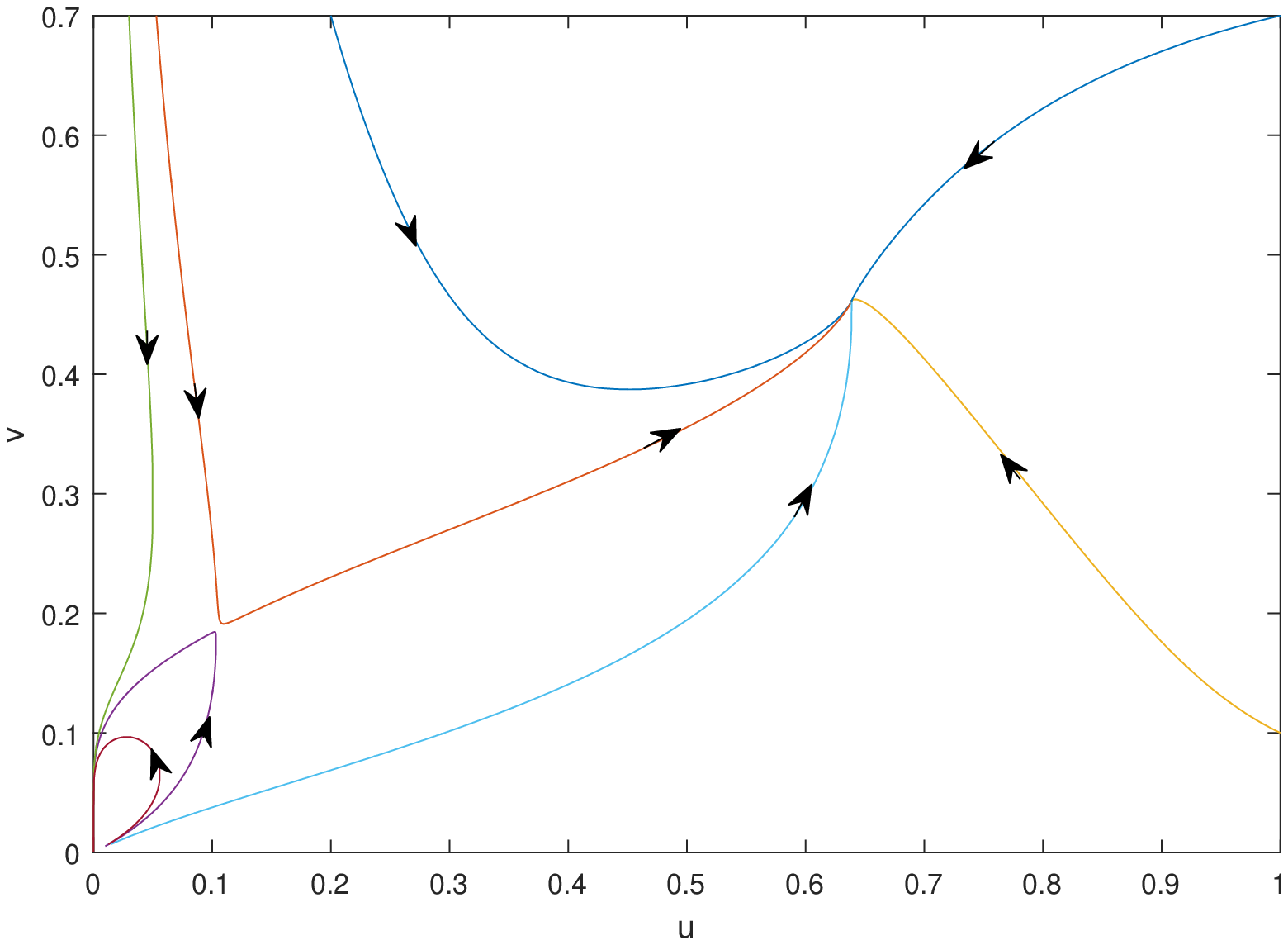}
\caption{Dynamics of (\ref{sm1})-(\ref{sm2}) when $r = 1/2, k = 1, H = 3, \nu = 2$.}
\label{F5}
\end{figure}

\begin{figure}[ht] 
\centering
\includegraphics[width=5.0in,height=4.0in]{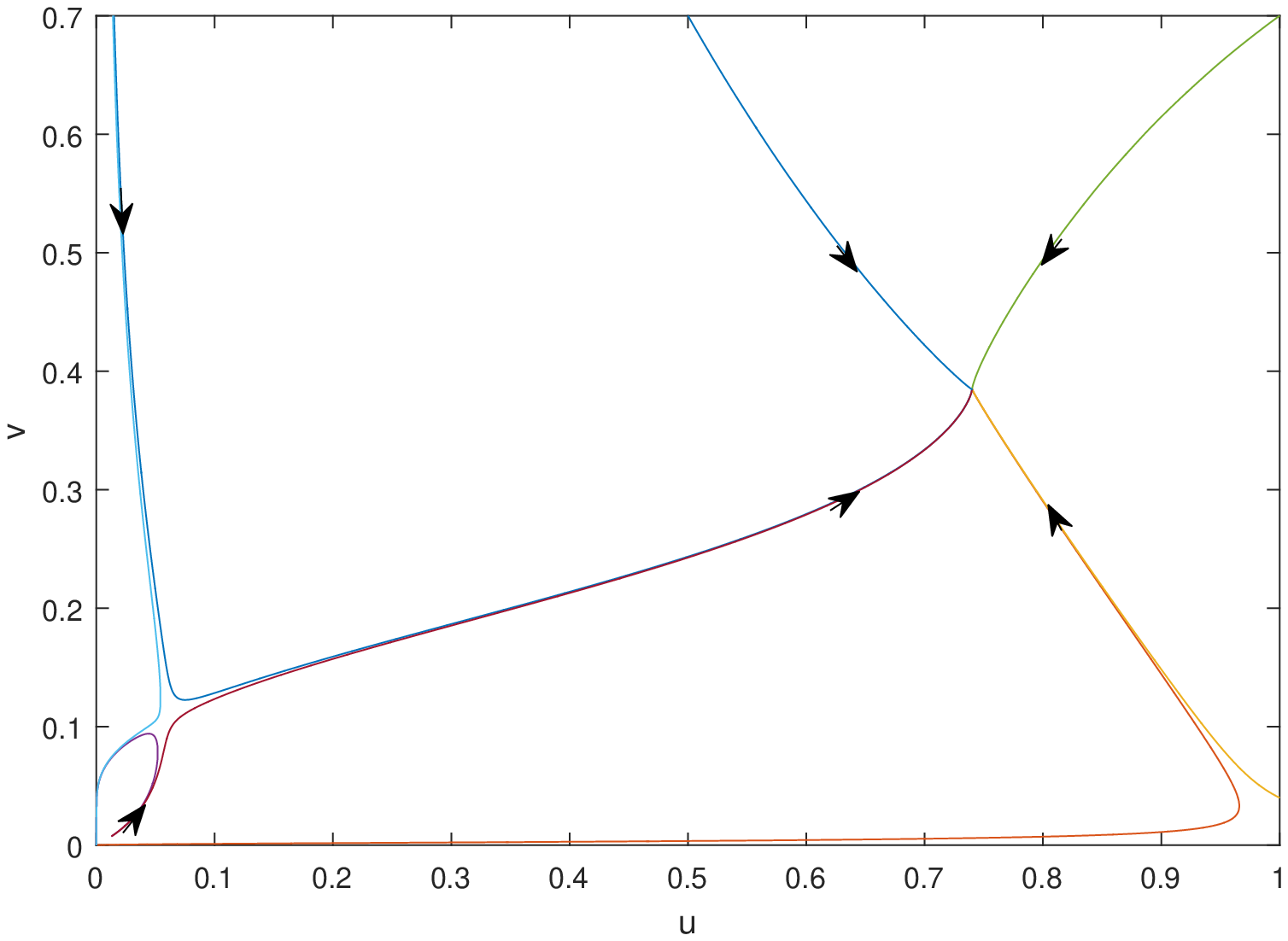}
\caption{Dynamics of (\ref{sm1})-(\ref{sm2}) when $r = 1/2, k = 1, H = 5, \nu = 2$.}
\label{F9}
\end{figure}

\begin{figure}[ht] 
\centering
\includegraphics[width=5.0in,height=4.0in]{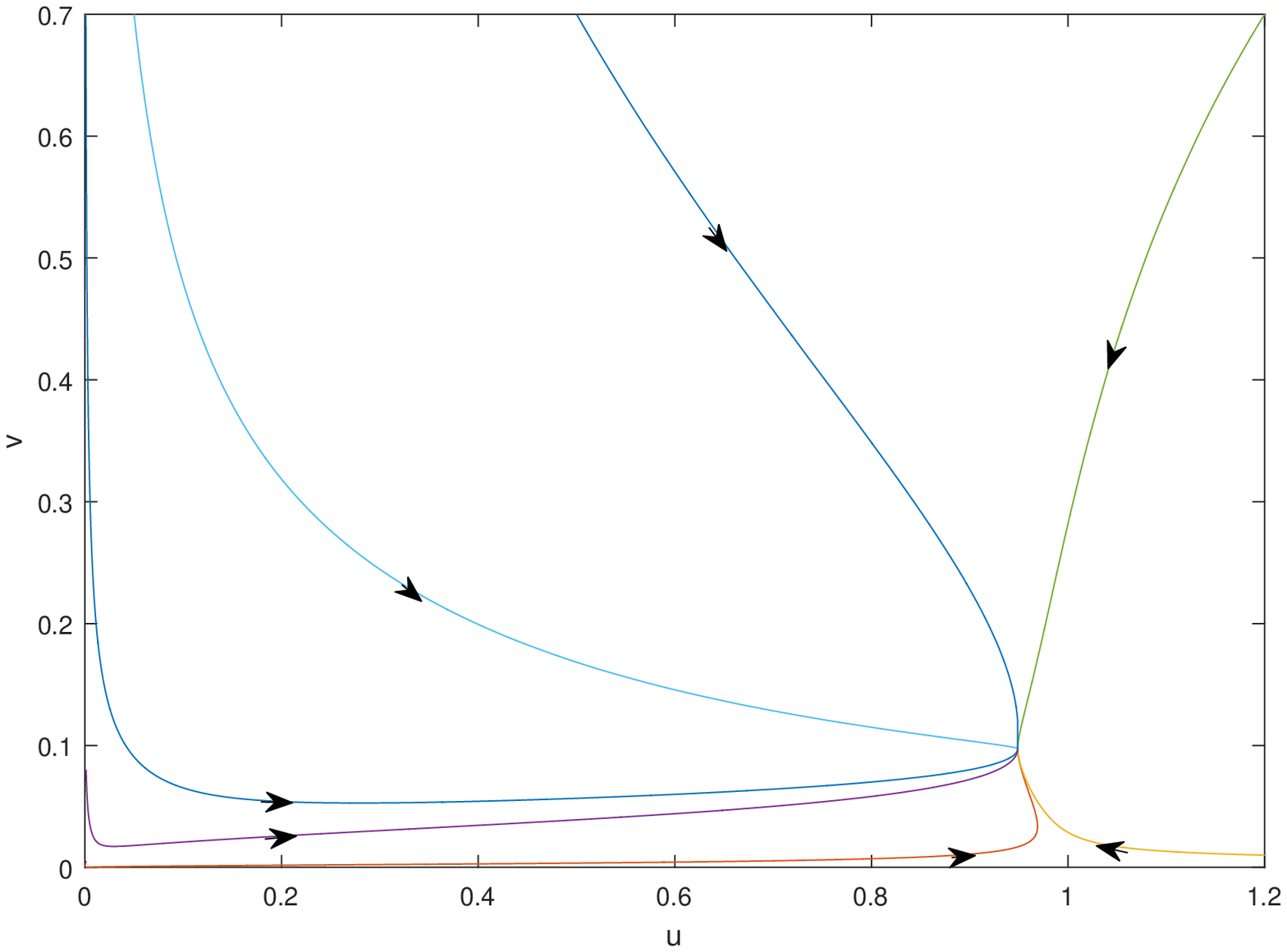}
\caption{Dynamics of (\ref{sm1})-(\ref{sm2}) when $r = 1/2, k = 1, H = 100, \nu = 2$.}
\label{F9b}
\end{figure}

\begin{figure}[ht] 
\centering
\includegraphics[width=5.0in,height=4.0in]{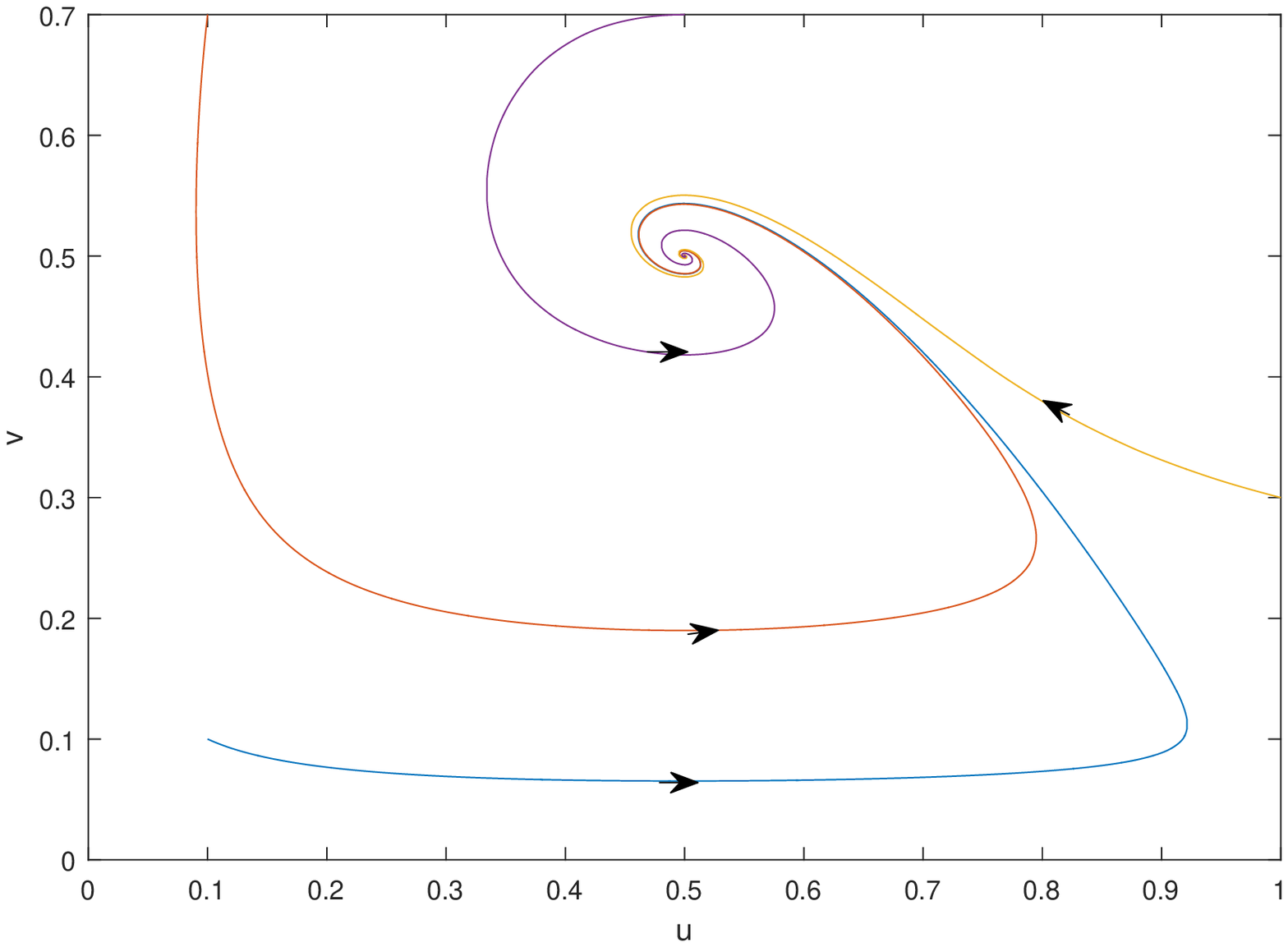}
\caption{Dynamics of (\ref{sm1})-(\ref{sm2}) when $r = 1/2, k = 1, H = 0.5, \nu = 1/2$.}
\label{F3}
\end{figure}

\begin{figure}[ht] 
\centering
\includegraphics[width=5.0in,height=4.0in]{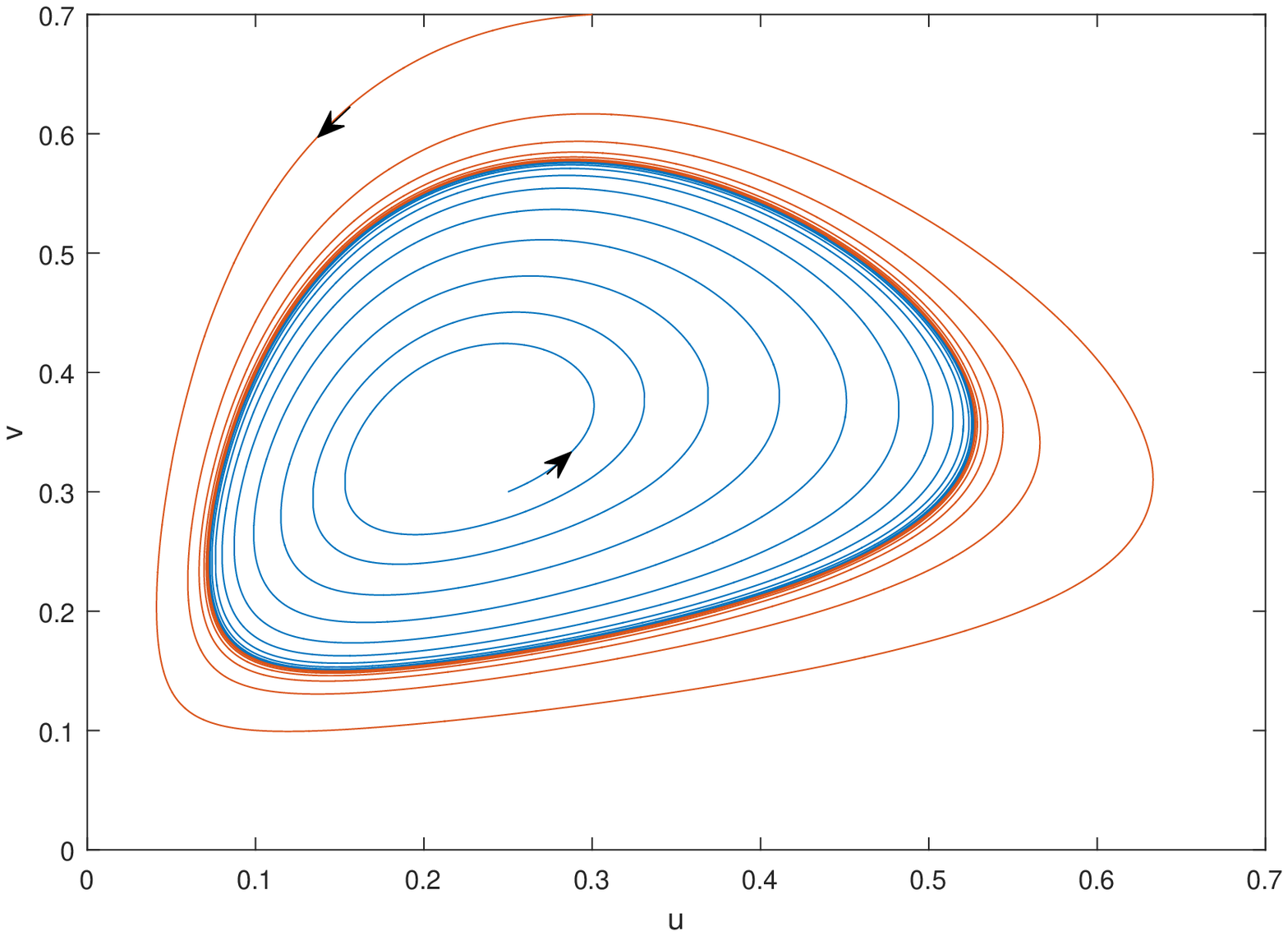}
\caption{Dynamics of (\ref{sm1})-(\ref{sm2}) when $r = 1/2, k = 1, H = 0.38, \nu = 1/2$.}
\label{F2}
\end{figure}

\begin{figure}[ht] 
\centering
\includegraphics[width=5.0in,height=4.0in]{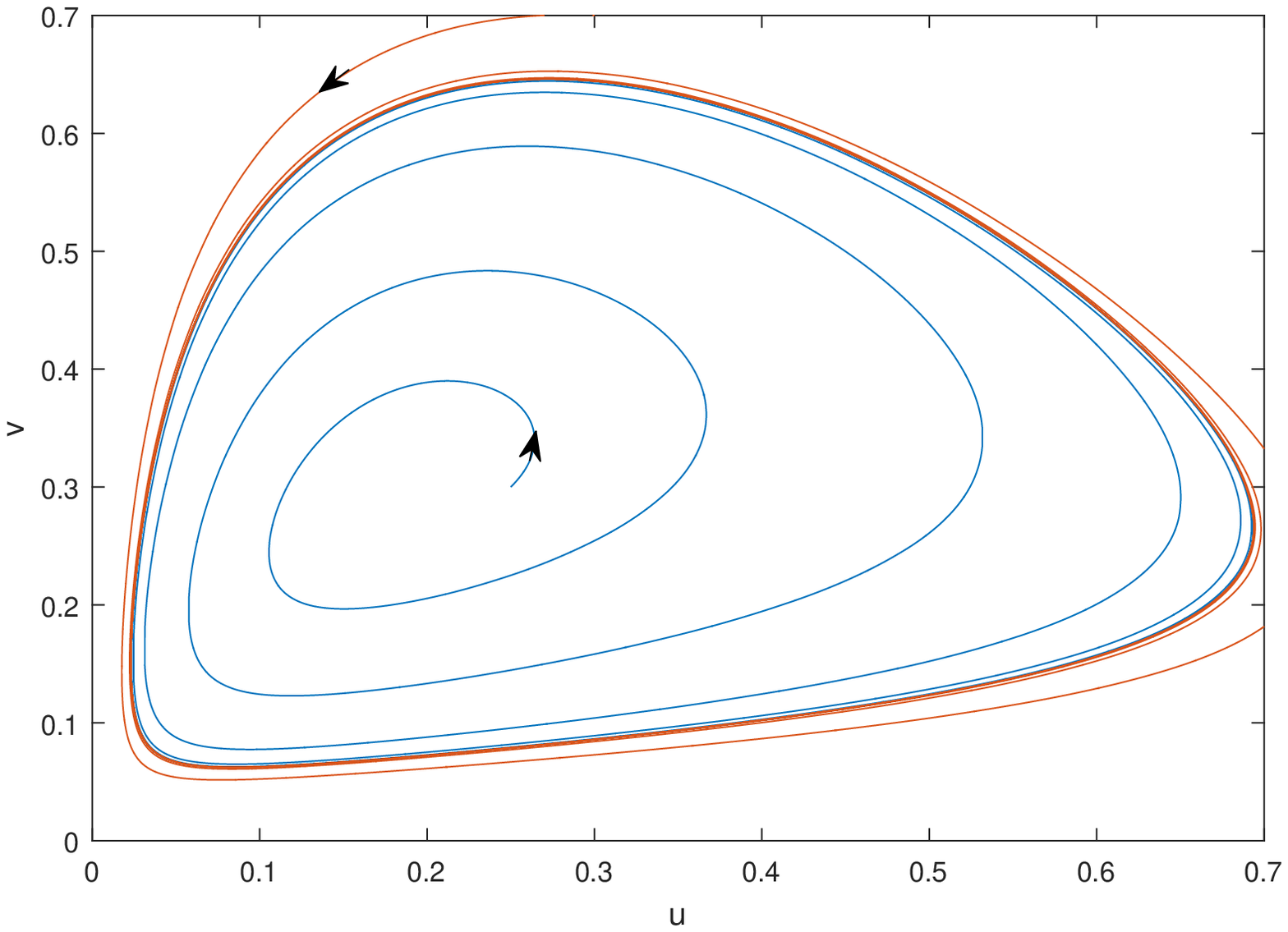}
\caption{Dynamics of (\ref{sm1})-(\ref{sm2}) when $r = 1/2, k = 1, H = 0.34, \nu = 1/2$.}
\label{F2b}
\end{figure}

We will be studying the following example of the general model (\ref{GM}),
\begin{eqnarray*}
\frac{dU}{dt} &=& \al U \left ( 1-\frac{U}{b} \right ) - \ga \frac{\frac{U}{V^{\nu}}}{\frac{U}{V^{\nu}} +h}V, \\  
\frac{dV}{dt} &=& \k \ga \frac{\frac{U}{V^{\nu}}}{\frac{U}{V^{\nu}} +h}V- \mu V,
\end{eqnarray*}
where the parameters are as given before. After introducing the dimensionless quantities and parameters: 
\begin{eqnarray*}
&& u = \frac{U}{b}, \  v = \frac{\ga}{\al b}V, \  \tau = \al t , \\
&& H = b^{\nu -1} h \left (\frac{\al}{\ga}\right)^{\nu}, \ k = \frac{\k \ga}{\al}, \  r = \frac{\mu}{\k \ga},
\end{eqnarray*}
we get the dimensionless form of the model
\begin{eqnarray}
\frac{du}{d\tau} &=& u( 1-u) - \frac{\frac{u}{v^{\nu}}}{\frac{u}{v^{\nu}} +H}v, \label{sm1} \\  
\frac{dv}{d\tau} &=& k \left (\frac{\frac{u}{v^{\nu}}}{\frac{u}{v^{\nu}} +H} - r \right ) v, \label{sm2}
\end{eqnarray}
where $H$ is named the capacity-predation number, and $r$ is named the mortality-food number. Enrichment corresponds to the increase of the carrying capacity $b$. Thus when $0 \leq \nu <1$, enrichment corresponds to the decrease of $H$, while 
when $\nu >1$, enrichment corresponds to the increase of $H$. When $\nu = 1$, enrichment does not change $H$.  When $\nu >0$, the term
\[
S(u,v) = \frac{\frac{u}{v^{\nu}}}{\frac{u}{v^{\nu}} +H}v
\]
on the right hand sides of (\ref{sm1})-(\ref{sm2}) is singular as $v \ra 0^+$. We do have the limit
\[
\lim_{u\ra u_*, v\ra 0^+} S(u,v) = 0 ,
\]
for any $u_* \geq 0$. The partial derivatives of $S$ are given by
\begin{eqnarray*}
&& \frac{\pa S}{\pa u} = H \frac{\frac{1}{v^{\nu}}}{\left (\frac{u}{v^{\nu}} +H\right )^2}v , \\
&&\frac{\pa S}{\pa v} =  \frac{\frac{u}{v^{\nu}}}{\frac{u}{v^{\nu}} +H} \left ( \frac{H}{\frac{u}{v^{\nu}} +H} + 1 \right ) .
\end{eqnarray*}
We have 
\[
\lim_{u\ra u_*, v\ra 0^+}  \frac{\pa S}{\pa u} = 0 , \  \lim_{u\ra u_*, v\ra 0^+}  \frac{\pa S}{\pa v} = 1 ,
\]
for any $u_* >0$. If $0<\nu <1$, 
\[
\lim_{u\ra 0^+, v\ra 0^+}  \frac{\pa S}{\pa u} = 0.
\]
If $\nu >1$,
\[
\lim_{u\ra 0^+, v\ra 0^+}  \frac{\pa S}{\pa u}
\]
does not exist. Finally we have 
\[
\lim_{u\ra 0^+, v\ra 0^+}  \frac{\pa S}{\pa v}
\]
does not exist for any $\nu >0$. Thus when $\nu >0$, ($u,v$)=($0,0$) and ($1,0$) are fixed points of the system (\ref{sm1})-(\ref{sm2}), and the Jacobian matrix of the right hand sides of the system (\ref{sm1})-(\ref{sm2}) does not exist at ($0,0$), and at ($1,0$) is given by
\[
\left ( \begin{array}{lr} -1 & -1 \cr 0 & k(1-r) \cr \end{array} \right ) .
\]
Thus, ($1,0$) is a saddle when $r < 1$, and a stable node when $r>1$. That is, when the mortality rate of the predator is high enough, the predator will go extinct, and the prey population will reach its full capacity, see Figure \ref{F1} for an illustration. When $r<1$, there may be other fixed points given by
\begin{equation}
v = \frac{1}{r} u(1-u), \  v = \left( \frac{1-r}{rH}\right)^{1/\nu} u^{1/\nu} . \label{ofp}
\end{equation}
For example, in the case of $\nu =1$ of Arditi-Ginzburg, when $H>1-r$, another fixed point is given by
\[
u = 1- \frac{1-r}{H}, \  v = \frac{1-r}{rH} \left ( 1 - \frac{1-r}{H} \right ).
\]
When $\nu >1$, there are two other fixed points when $H$ is large enough, otherwise there is no more fixed point. Enrichment here corresponds to increasing $H$. When $H$ is not large enough, there is no extra fixed point, the dynamics always ends up 
extinction of both the predator and the prey ($0,0$), see Figure \ref{F4} for an illustration. Increasing $H$ to some value, the two extra fixed points are about to emerge as shown in Figure \ref{F5c}. The  two extra fixed points emerge via saddle-node bifurcation at 
\[
u_c = \frac{\nu - 1}{2\nu - 1}, \  v_c=\frac{1}{r} u_c (1-u_c), \  H_c = \frac{1-r}{r}  \frac{u_c}{v_c^{\nu}} ,
\]
where the two curves in (\ref{ofp}) tangentially intersect at ($u_c,v_c$). For example, when $r=1/2$ and $\nu = 2$, the critical $H$ 
value should be $H_c = 1.6875$ which is very close to the $H$ value in  Figure \ref{F5b}.
After the two extra fixed points emerge, one of them is stable, 
while the other is a saddle as shown in Figures \ref{F5b} and \ref{F5}. Increasing $H$ further, the stable fixed point approaches ($1,0$) as shown in Figures \ref{F9} and \ref{F9b}, i.e. extinction of the predator and the prey population reaches its capacity. Thus 
when $\nu >1$, low carrying capacity of prey causes extinction of both prey and predator. Moderate enrichment can avoid extinction and lead to stable prey and predator populations. Extreme enrichment can lead to the extinction of the predator, while the prey 
population reaches its capacity. When $0<\nu < 1$, there is always another fixed point. For example, in the case $\nu = 1/2$, the extra fixed point is given by
\[
u = \left [ 1 +\frac{1}{r} \left (\frac{1-r}{H}\right )^2 \right ]^{-1}, \ 
v = \left (\frac{1-r}{rH}\right )^2 \left [ 1 +\frac{1}{r} \left (\frac{1-r}{H}\right )^2 \right ]^{-2}.
\]
When $\nu = 0$, there is another fixed point when $H<\frac{1}{r} - 1$ as shown in \cite{FL15}. When $0\leq \nu <1$, enrichment corresponds to decreasing $H$. When the carrying capacity of the prey is not large enough, the extra fixed point is stable as 
shown in Figure \ref{F3}, there is no extinction, and the dynamics always ends up at stable prey and predator populations given by the stable fixed point. Enrichment will cause the stable fixed point to lose it stability, and a stable limit cycle will emerge around it 
as shown in Figure \ref{F2}. Further enrichment will cause the limit cycle to get closer and closer to predator and prey axes  as shown in Figure \ref{F2b}, and in such a situation, environmental fluctuation can cause extinction of the predator and the prey. Thus, 
enrichment always leads to extinction in both the cases $\nu >1$ and the cases $0 \leq \nu <1$, except 
the case $\nu =1$. When $\nu >1$, enrichment reduces the predation intensity, and leads to extinction of predator. When $0 \leq \nu <1$, enrichment increases the predation intensity, and leads to extinction of prey together with predator. In summary, our mathematical model here almost always predicts the existence of the enrichment paradox. 


\section{Model applications}

It is very difficult to verify any model in real field predator-prey systems not only because collecting and testing real field data against a model is a daunting job, but also because real field predator-prey systems are rarely binary (i.e. isolated systems of a 
predator population and a prey population), rather consisting of many populations forming a food chain or even a food web. Here we will attempt rather qualitative applications of the model.

\subsection{Model application to human-food dynamics}

In human history, technological advances are the main factors for human population increase, such as tool-making revolution, agricultural revolution and industrial revolution. Technological advances provide human with more food supply and medicine. 
Diseases such as plagues  also caused human population to temporarily decrease. But since 1700, human population has been monotonically increasing due to technological advances. Since 1960s, due to the introduction of high yield grains, agricultural 
machineries, fertilizers, chemical pesticides, and better irrigation systems, human population has been increasing by 1 billion every 12 years, from 3 billion to 8 billion by 2023. Thus enrichment of food has increased human population dramatically. It seems that both the Cornucopian and the Malthusian views were realized \cite{Kha77} \cite{Mal26}. Human indeed dramatically advanced technology to provide abundant food supply to meet the demand of population growth according to Cornucopian view. Human population also dramatically increased with the abundant food supply according to Malthusian view. The question is whether or not we are heading to a new Malthusian catastrophe, i.e. some people are going to starve. Technologies may be advanced further to support more humans. But the earth 
resource is limited, and the human population cannot increase unlimited on earth. Is the human population following a path to a stable steady state as in Figure \ref{F5}? or to a disaster as in Figure \ref{F4}? or a less severe disaster as in Figure \ref{F2}? Human 
overpopulation not only can cause huge damage to earth resource and environment, but also has serious sustainability consequence. 
If there is a global food scarcity, huge famine can cause major population loss. According to World Wide Fund for Nature \cite{WWF06}, the current human population is already exceeding its earth carrying capacity. On the other hand, estimating earth's carrying capacity for human is more difficult than for other animals due to the fact that human choices may play an important role \cite{Coh95}. In the long run, human population cannot continue to grow. There are clear human resource limits of food, energy and territory (individual human space) as discussed by von Hoerner \cite{Hoe75}. The key moment is when human population reaches its maximum. The crucial question is: How will the human population change afterward? Will human population more or less stay at a stagnation population or decrease substantially? If human population 
decreases, is the decrease due to birth control, normal death or abnormal death? Birth control and normal death are hopeful for reducing human population from the example of China. Abnormal death corresponds to various kinds of disasters such as diseases, wars etc.. Von Hoerner also proposed the possibility of moving humans out of earth, i.e. stellar expansion \cite{Hoe75}. But wars and diseases are more probable. 

There have been a lot of efforts in fitting human population historical data with a function such as the nice fitting by $c(t_* - t)^{-\al}$ for positive
 parameters $c$, $t_*$ and $\al$ 
\cite{KM16}. But human population growth is a complex dynamics of an extremely complex system, which is extremely difficult to predict.

\subsection{Model application to amoebae population - life finds a way}

Amoebae-food system can also be viewed as a predator-prey system. The dynamics of amoebae-food system is like that shown in 
Figure \ref{F4}. The amoebae population will continue to multiply until food is scarce. Then they will aggragate and build stem and fruiting structure. Majority of the amoebae will die, and only the amoebae inside the fruits will survive. When the next round of food arrives, the fruits will fall from the tops of the stems, and the amoebae inside fruits will come out and start the next round of multiplying game. So each round of the multiplying will end up with eradication of the majority of the amoebae. But life finds a way for amoebae species to survive - through fruiting and sacrifice of the majority of their population. 

\subsection{Human - food vs. amoebae - food dynamics}

Amoebae are single-celled organisms, and they have no control on their population growth when the food is abundant. With the increase of food supply, humans did not have a good control or plan on their population growth either. Facing scarcity of food, amoebae take the approach of sacrifice to ensure the survival of their species. 
Four thousand years ago, during the Yao and Shun dynasties in China, old people would starve to death to ensure the survival of young people when there was food shortage. Which approach will humans take when there will be a global food shortage? Will humans take the approach of amoebae? Amoebae have no brain, but so far 
humans did not plan their survival better than amoebae. There are a lot similarities in the population development between amoebae and humans. Studying amoebae can be important in understanding human life and survival.

\subsection{Model application to salmon farming}

Human-salmon system is a predator-prey system. Salmon farm can be viewed as an enrichment to salmon population. 
The population of human eating salmon and the population of salmon follow the dynamics like in Figure \ref{F2}. When the 
salmons are abundant, the price of salmons (especially farm raised salmons) will drop, and that will lead to more people to consume 
salmons. In return, the salmon supply will drop, lead to higher price, and less people will consume salmons. Due to health problems associated with salmon farming, people prefer wild salmon, and that caused severe predation on wild salmons, and led to the extinction of some wild salmons such as the Atlantic salmons.

\end{document}